\documentclass{jkas}

\def\year{2022} 
\def\month{Aug} 
\def\volume{00} 
\def\issue{0} 
\def\beginpage{1} 
\def\endpage{8} 
\def\accepted{July 25, 2021} 
\date{accepted \accepted}

\usepackage{flushend} 
\usepackage[utf8x]{inputenc} 
\usepackage{graphicx, natbib, romannum, verbatim, amsmath, caption}

\title{Newly discovered $z\sim5$ quasars based on deep learning and Bayesian information criterion}

\author[1]{Suhyun Shin}
\author[1]{Myungshin Im}
\author[2,3]{Yongjung Kim}
\author[3]{Linhua Jiang}

\affil[1]{SNU Astronomy Research Center, Astronomy Program, Dept. of Physics \& Astronomy, Seoul National University, 1 Gwanak-ro, Gwanak-gu, Seoul 08826, Republic of Korea; \email{suhyun.shin.s2@gmail.com, myungshin.im@gmail.com}}
\affil[2]{Department of Astronomy and Atmospheric Sciences, College of Natural Sciences, Kyungpook National University, Daegu 41566, Republic of Korea}
\affil[3]{Kavli Institute for Astronomy and Astrophysics, Peking University, Beijing 100871, People{'}s Republic of China}

\def\corrauthor{Myungshin Im}
\def\runningauthor{Shin et al.}
\def\runningtitle{Newly discovered $z\sim5$ quasars}

\def\keywords{galaxies: active, galaxies: quasars: general, methods: data analysis, methods: observational, techniques: spectroscopic}

\def\abstracttext{We report the discovery of four quasars with $M_{1450} \gtrsim -25.0$ mag at $z\sim5$ and supermassive black hole mass measurement for one of the quasars. They were selected as promising high-redshift quasar candidates via deep learning and Bayesian information criterion, which are expected to be effective in discriminating quasars from the late-type stars and high-redshift galaxies. The candidates were observed by the Double Spectrograph on the Palomar 200-inch Hale Telescope. They show clear Ly$\alpha$ breaks at about 7000-8000 \r{A}, indicating they are quasars at $4.7 < z < 5.6$. For HSC J233107-001014, we measure the mass of its supermassive black hole (SMBH) using its C\Romannum{4}$\lambda 1549$ emission line. The SMBH mass and Eddington ratio of the quasar are found to be $\sim 10^8 M_{\odot}$ and $\sim 0.6$, respectively. This suggests that this quasar possibly harbors a fast growing SMBH near the Eddington limit despite its faintness ($L_{\rm Bol} < 10^{46}$ erg s$^{-1}$). Our 100 $\%$ quasar identification rate supports high efficiency of our deep learning and Bayesian information criterion selection method, which can be applied to future surveys to increase high-redshift quasar sample.}

\begin{document}
\jkashead{}

\section{Introduction\label{sec:intro}}
Quasars are one of the most powerful astrophysical sources in the universe with a bolometric luminosity, $L_{\rm Bol} \gtrsim 10^{45}$ erg s$^{-1}$ (e.g., \citealp{Hickox+2018}). Owing to their high luminosities, we can find them even in the distant universe ($z>7$; \citealp{Mortlock+2011, Banados+2018, Matsuoka+2019, Yang+2020a, Wang+2021}). The discovery of quasars in the early universe enables us to investigate how the intergalactic medium (IGM) was ionized and supermassive black holes (SMBHs) grew (e.g.,  \citealt{Fan+2006, Trakhtenbrot+2011, KimYJ+2018, Onoue+2019, Yang+2020b}). By constructing the quasar luminosity function (LF) at $z\gtrsim4$ with several reasonable assumptions, the contribution of quasars to keep the ionized state of the IGM can be estimated \citep{Jiang+2016, Akiyama+2018, Matsuoka+2018b, McGreer+2018, Kim+2020, Shin+2020}. Given the short time available for SMBHs to grow between the birth of seed BHs and the age of the universe where the highest redshift known quasars reside, examining the properties of the early universe quasars can give us insights on how the SMBH became to exist (e.g., \citealt{Banados+2018, Yang+2020a}).

\indent Although the expected contribution to the UV background at $z\sim5$ of quasars with the rest-frame absolute magnitude at 1450 \r{A} ($M_{1450}$) between $-25$ and $-22$ mags are comparable to or greater than that of brighter quasars \citep{Kim+2020}, the number of the currently identified faint quasars with $-25 < M_{1450} < -22$ is much less than the number of brighter quasars so far. This is because previous quasar searches have mostly relied on wide-field surveys with shallow depths. The situation is improving with deeper data becoming available for high-redshift quasar search: the Canada–France–Hawaii Telescope Legacy Survey (CFHTLS; \citealt{Gwyn2012}); the Dark Energy Survey (DES; Dark Energy Survey Collaboration et al. 2016); the Hyper Suprime-Cam Subaru Strategic Program (HSC-SSP; \citealt{Aihara+2019}); the Infrared Medium-deep Survey (IMS; M. Im et al. 2022, in preparation).

\defcitealias{Niida+2020}{N20}
\indent However, finding faint quasars is more challenging than bright quasar search, even with deeper images. High-redshift quasars can be identified by the strong break in their spectral energy distributions (SEDs) caused by the redshifted Lyman break \citep{Jiang+2016, Jeon+2017, McGreer+2018, KimYJ+2019, Shin+2020}. However, as we go fainter in magnitude, there is a rapid increase in the number of other types of objects, such as primarily late-type stars and high redshift galaxies, which mimic this break. Conventional color selection considering only a few broadband colors is not enough to reject lots of contaminant sources in faint quasar candidates (\citealt{Matsuoka+2018a, Niida+2020}, hereafter, \citetalias{Niida+2020}). SED model fitting is a physically meaningful approach for searching quasars \citep{Reed+2017}. However, it would require a considerable amount of computing resources if one hopes to find faint quasars among many contaminants.

\indent In our previous work (\citealp{Shin+2022}, accepted), we developed a novel method for selecting quasar candidates adopting the deep learning and the Bayesian information criterion (BIC). We applied this method to the HSC-SSP data that reaches a 5-$\sigma$ depth of $i\sim26$ mag for point source detection \citep{Aihara+2019} which corresponds to $M_{1450} \sim -20$ mag for $z\sim5$ quasars. We identified 35 faint quasar candidates, five being previously known quasars.

\indent In this paper, we report our spectroscopic observation of four quasar candidates with $i < 23$ mag as an attempt to further confirm the effectiveness of our new quasar selection method. In Section~\ref{sec:selection}, we describe the photometric selection process for quasars at z $\sim5$ in brief. The specification for the spectroscopic observation is explained in Section~\ref{sec:specinfo}. Section~\ref{sec:spectralfitting} includes the spectral fitting procedure and black hole (BH) mass measurement from C\Romannum{4} emission line. The efficiency of our quasar selection is addressed in Section~\ref{sec:efficiency}. Section~\ref{sec:sum} summarizes our findings. We assume cosmological parameters $\Omega_{M} = 0.3$, $\Omega_{\Lambda} = 0.7$ and $H_{0} = 70$ km~s$^{-1}$ Mpc$^{-1}$ throughout the paper. We adopted the AB magnitude system for representing a flux measured in each filter \citep{Oke+1983} and used the dust map of \citet{Schlegel+1998} to correct fluxes for the Galactic extinction.

\begin{table*} 
\caption{General information of the targets and spectroscopic observations\label{tab:info}}
\centering
\begin{tabular}{llrrrrrrrll}
\toprule 
ra & dec & $g$ & $r$ & $i$ & $NB816$ & $z$ & $NB921$ & $y$ & Date & ExpTime \\
(J2000) & (J2000) & [mag] & [mag] & [mag] & [mag] & [mag] & [mag] & [mag] & & [sec] \\
\midrule
02:17:33.44 & -4:44:44.32 & 25.7 & 23.7 & 22.2 & 22.4 & 22.1 & 22.1 & 22.3 & 2020 Nov 19 & 1500 \\ 
16:18:27.28 & 55:17:48.51 & 25.3 & 22.7 & 21.1 & 21.1 & 20.9 & 20.9 & 20.8 & 2021 Jul 13 & 600 \\ 
23:27:13.22 & 0:05:47.92 & $>$ 27.3 & 25.1 & 22.3 & 21.3 & 22.0 & 21.7 & 21.8 & 2020 Nov 19 & 1800 \\ 
23:31:07.00 & -0:10:14.52 & $>$ 27.3 & 24.4 & 22.6 & 23.2 & 22.6 & 21.8 & 22.7 & 2020 Nov 19 & 1800 \\ 
\bottomrule
\end{tabular}
\tabnote{The magnitude errors are mostly less than 0.03 mag.}
\end{table*}

\section{Quasar selection \label{sec:selection}}
Faint quasar candidates at $z\sim5$ were selected in our previous work (\citealp{Shin+2022}, accepted). Here, we briefly explain how the quasar candidates were selected.

\subsection{Photometric data}
The HSC-SSP has Wide, Deep, and UltraDeep layers. Taking advantage of deep images ($i \sim 26$ mag) and moderate survey area (27 deg$^{2}$), we choose the Deep layer to search quasars among the layers. The Deep layer consists of four fields covered by five broadbands ($g, r, i, z, y$) and two narrow-bands ($NB816, NB921$). The 5-$\sigma$ image depths of broadbands ($g, r, i, z, y$) are (27.3, 26.9, 26.7, 26.3, and 25.3). For $NB816$ and $NB921$, the depths are 26.1 and 25.9 mag, respectively. 

\indent We retrieved a source catalog of the second public data release (PDR2) of the HSC-SSP. To exclude objects with unreliable photometry, we adopted flags evaluating influences from the bad pixel, cosmic ray, saturation at the center of an object, abnormal background level, and the location of an object on an image. Also, we constrained that a source should be a primary object with detection in the $i$-band. After excluding the flagged objects, the resulting effective survey area of the Deep layer for our quasar search became $\sim 15.5$ deg$^{2}$ based on a random source catalog in HSC-SSP data archival system \citep{Coupon+2018}, somewhat smaller than the nominal HSC-SSP Deep survey area of 27 deg$^2$. The number of sources in the catalog is 3.5 million. In this study, we use two magnitude systems: the point-spread function (PSF) magnitude and the CModel magnitude (CModel). 

\subsection{SED models} \label{sec:SEDmodels}
Due to the small sample size of spectroscopically confirmed quasars at $z\sim5$ in the Deep layer of the HSC-SSP ($<10$), quasar SED models were used for training deep learning models and calculating BIC. Stellar SED models are also necessary to do SED fitting. Since these SED models were introduced in detail in \citealt{Shin+2022}, accepted, here we explained them in brief.

Our quasar SED models have four free parameters: redshift ($z_{\rm SED}$), continuum slope ($\alpha_{\lambda}$), equivalent width of Ly$\alpha$ and N \Romannum{5} $\lambda 1240$ (EW), and $M_{1450}$. We created a composite SED by adopting the SED at $\lambda < 1450$ \r{A} from \citet{Lusso+2015} and the redder part from \citet{Selsing+2016}. Then, the Ly$\alpha$+N V equivalent width and the continuum slopes are adjusted. The IGM attenuation model of \citet{Inoue+2014} was adopted.

For the stellar SED models, we used the BT-Settl models \citep{Allard+2013} which have five free parameters: effective temperature ($T_{\text{eff}}$), surface gravity (log($g$)), metallicity ([M/H]), alpha-element enrichment ([$\alpha$/M]), and a normalization factor (f$_{N}$).

\subsection{Quasar selection process}
\begin{figure}[t]
\centering
\includegraphics[width=85mm]{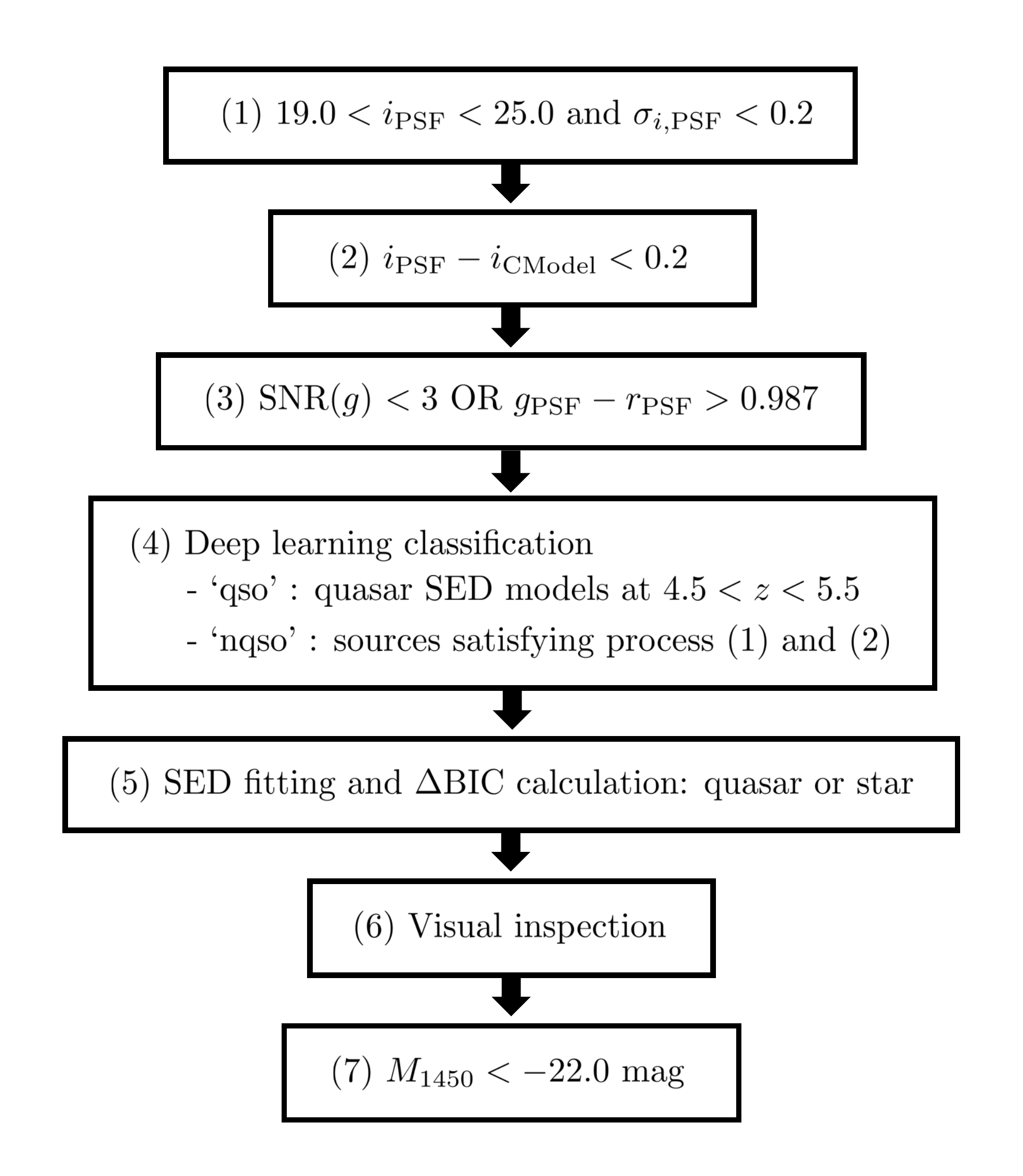}
\caption{Flow chart for the entire quasar selection process. \label{fig:flowchart}}
\end{figure}

We applied multiple criteria sequentially. Figure~\ref{fig:flowchart} shows the sequence of our selection. 

\indent First, we selected objects with $19 < i_{\rm PSF} < 25$ and magnitude error $<0.2$ mag (the process (1)). Then, we eliminated extended sources by applying the extendedness cut of $i_{\rm PSF} - i_{\rm CModel} < 0.2$ (the process (2)). With this procedure, we could include about 98.4 $\%$ of point sources identified by an $I$-band catalog of the Hubble Space Telescope (HST) Advanced Camera for Surveys \citep{Leauthaud+2007}. The number of candidates satisfying the process (1) and (2) was 333,780. Among them, there are six spectroscopically confirmed quasars at $4.5 < z < 5.5$ \citep{McGreer+2013, Paris+2018, Shin+2020}.

\indent Since faint quasars at $z\sim5$ have the strong IGM absorption lines in the blueward of the Ly$\alpha$ emission line, they tend to show a weak signal to noise ratio (SNR $< 3$) in the $g$-band or large $g-r$ color. To focus on the quasars at $4.5 < z < 5.5$, we limited the $g-r$ color to 0.987, which is the minimum $g-r$ value for quasar models at the redshift range. The number of candidates was 125,644. 

\indent After choosing the red objects, we performed classification using deep learning. We trained 100 models to predict a class of an object based on its HSC-SSP photometry information. Our trained models assume that the red objects belong to one of two classes: quasars at $z\sim5$ (‘qso’) or non-quasar sources (‘nqso’). Training set for the ‘qso’ comprised of 100,000 randomly sampled quasar models at $4.5 < z < 5.5$. For training ‘nqso’ class, 100,000 randomly sampled point sources with $19 < i < 25$ and $\sigma_{i} < 0.2$ were used. Combining results of 100 deep learning models, we achieved an average accuracy larger than 99 $\%$ for the ‘qso’ class. 1,599 candidates were selected in our ensemble learning.

\indent To compensate for the simple approximation used in the deep learning and deal with possible misclassified ‘nqso’ objects, we carried out SED fitting of 1,599 deep-learning-selected candidates using quasar and stellar SED models and compare the best-fit models by adopting the BIC. For each best-fit model, the BIC value can be calculated as,

\begin{equation}
\label{equ:BIC}
\rm{BIC} = \chi^2 + k \times \ln n,
\end{equation}

\noindent where $k$ is the number of free parameters in the model, $n$ is the number of the data, and $\chi^2$ is the chi-square value for the model.
The BIC imposes the penalty term to the model with a large $k$, so a fair comparison between given models becomes possible. The difference between the BIC values, $\Delta\rm{BIC}$, is defined as

\begin{equation}
\Delta\rm{BIC} = \rm{BIC}_{\rm{star}} - \rm{BIC}_{\rm{quasar}}.
\end{equation}

The larger $\Delta\rm{BIC}$ is, the more likely it is a quasar. We considered the minimum value of the $\Delta\rm{BIC}$ as 10 \citep{Liddle+2007}. The number of the BIC-selected candidates was 78.

\indent We visually inspected multi-band HSC-SSP images to remove candidates for which the photometry could be affected by satellite tracks, nearby bright stars, optical ghosts, and scattered lights. Among the 78 BIC-selected candidates, 53 candidates passed the visual inspection.

\indent Finally, we restricted the $M_{1450}$ of the candidate to be brighter than -22.0 mag where the number density of quasar candidates can dramatically increase due to the galaxy contamination (\citetalias{Niida+2020}, \citealt{Shin+2022}, accepted). The number of the final quasar candidates is 35. Five out of 35 candidates are known quasars at $z\sim5$ \citep{McGreer+2013, Shin+2020}. Another four candidates were selected to be promising candidates selected in \citet{McGreer+2018}, \citet{Chaves-Montero+2017} and \citet{Shin+2020}. Among these four candidates, a spectrum of one candidate \citep{Shin+2020} was obtained in this study.

\begin{table*} 
\caption{Best-fit spectral parameters\label{tab:specfit}}
\centering
\begin{tabular}{lrrrrr}
\toprule 
id & $z_{\mathrm{Ly}\alpha}$ & $z_{\mathrm{spec}}$ & $\alpha_{\lambda}$ & $\log (EW)$ & $M_{1450}$ [mag] \\
\midrule
HSC J021733-044444 & 4.807 & $4.783_{-0.003}^{+0.003}$ & $-1.58_{-0.37}^{+0.35}$ & $2.38_{-0.05}^{+0.05}$ & $-23.88_{-0.06}^{+0.06}$ \\ 
HSC J161827+551748 & 4.754 & $4.726_{-0.008}^{+0.005}$ & $-1.94_{-0.56}^{+0.53}$ & $1.71_{-0.17}^{+0.12}$ & $-25.13_{-0.06}^{+0.06}$ \\ 
HSC J232713+000547 & 5.591 & $5.561_{-0.006}^{+0.003}$ & $-1.40_{-0.70}^{+0.74}$ & $2.51_{-0.06}^{+0.06}$ & $-24.44_{-0.09}^{+0.10}$ \\ 
HSC J233107-001014 & 4.974 & $4.950_{-0.004}^{+0.002}$ & $-0.58_{-0.56}^{+0.52}$ & $2.62_{-0.10}^{+0.10}$ & $-23.03_{-0.16}^{+0.19}$ \\ 
\bottomrule
\end{tabular}
\end{table*}

\section{Spectroscopy \label{sec:specinfo}}
We conducted spectroscopic observations for four candidates with $i_{\rm PSF} < 23$ mag between 2020 November and 2021 July. Note that one of the targets, IMS J161827+551748 was also selected using a different selection method that included medium-band data \citep{Shin+2020}. We utilized the Double Spectrograph (DBSP) on the 200-inch Hale Telescope in the Palomar Observatory (PID: CTAP2020-B0043 and CTAP2021-A0032, PI: Y.Kim). Using the dichroic filter, the DBSP can observe the red and blue channels simultaneously. The dichroic filter we used was D55. We set a long-slit mode with a slit of which the width and length correspond to 1.5 and 128 arcseconds. The 316 (600) lines$/$mm grating with a wavelength of 7500 (4000) \r{A} was chosen for the red (blue) channel. The exposure time was $\sim$ 1200 to 1800 seconds. The typical seeing during the observing period was $\sim$ 1.1 to 1.5 arcseconds. 

\indent After acquiring the spectroscopic data, we pre-processed the data with a python package, {\tt \string PypeIt} \citep{Prochaska+2020a, Prochaska+2020b}. Owing to the expected IGM absorption in the blue channel, we only considered the red channel in this study. The {\tt \string PypeIt} can automatically subtract bias, perform flat-fielding, give a wavelength solution, and model the sky background. To do wavelength calibration, we used a HeNeAr lamp. Since our targets are too faint to be detected with a default algorithm, we manually extracted the fluxes of a standard star and the targets. We adjusted the spatial and spectral locations of the extraction window and its size to maximize the signal-to-noise ratio of each spectrum.

\indent After reducing the data, we calculated a sensitivity function by comparing the observed fluxes of a standard star, Feige110, with its actual fluxes. After the sensitivity correction, we rescaled the spectrum using HSC-SSP $i$-band magnitude to correct for a possible flux loss due to the finite size of the slit width.

\indent From the full width at half maximum (FWHM) of sky emission lines, we estimated a spectral resolution, $R = 800 - 1300$, which varied depending on the observed wavelength and the observation date. Table~\ref{tab:info} shows the coordinate, HSC-SSP photometry, observation date, and exposure time for the candidates.  

\begin{figure}[t]
\centering
\includegraphics[width=85mm]{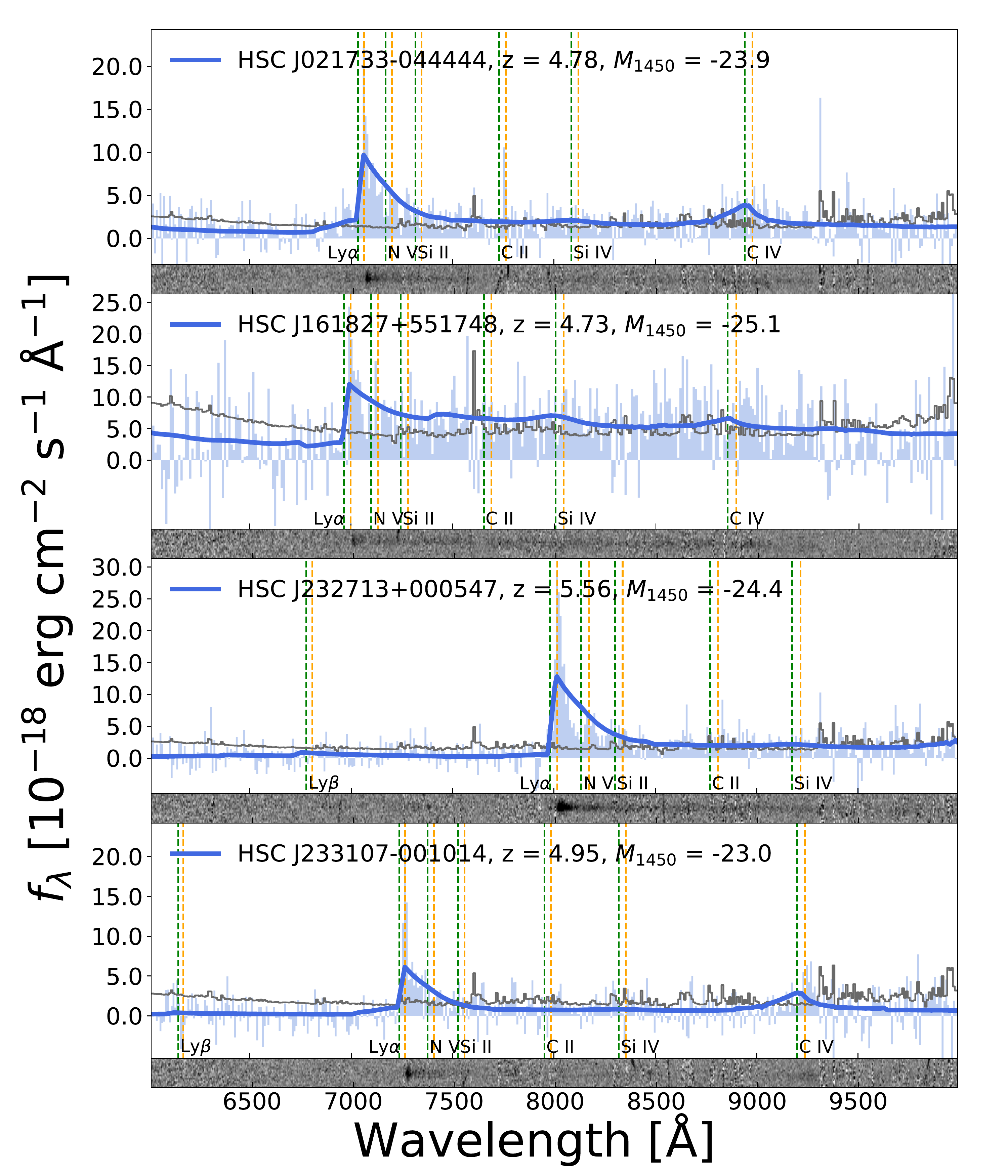}
\caption{The 1-d and 2-d spectra of four observed quasar candidates in sequence. The skyblue bins and gray lines correspond to the 1-d binned spectra and their 1-$\sigma$ flux errors, respectively. The blue thick lines are the best-fit quasar SED model. Several dominant lines in a typical quasar spectrum are marked as green (orange) vertical lines of which the locations depend on $z_{\mathrm{spec}}$ ($z_{\mathrm{Ly}\alpha}$). The 2-d spectra are shown in an inverted gray scale. \label{fig:qsospec}}
\end{figure}


\section{Result \label{sec:spectralfitting}}

\subsection{Spectral fitting}
To increase the SNR of each spectrum, the spectrum was binned at intervals of $6\sim9$ \r{A} ($5\sim6$ pixels), corresponding to FWHM from its $R$. Each flux in each bin was weighted by the inverse of its squared flux uncertainty. The weighted mean flux is to be a representative value of each bin. After binning, the SNRs of the spectra reach 5 to 15 at their emission lines, and 2 to 3 at their continua. Figure~\ref{fig:qsospec} shows the binned spectra of the four objects. All four objects show predominant and broad Ly$\alpha$ emission lines with sufficient $SNR\sim10$ and sharp breaks at the blueward of the line due to the IGM absorption, confirming their nature as high-redshift quasars. 



\indent To estimate the spectroscopic redshift ($z_{\mathrm{spec}}$), we performed a spectrum fitting based on quasar SED models to the binned spectra (please refer to Section~\ref{sec:SEDmodels}). The best-fit model was obtained by using Markov Chain Monte Carlo (MCMC) method \citep{emcee+2013}. From the sampled posterior distribution of the parameter, we obtained the best-fit parameter and its error as the median and the 68$\%$ equal-tailed interval of the distribution (Table~\ref{tab:specfit}). Due to the low SNR of the continuum, the uncertainties of the parameters are somewhat large except for $z_{\mathrm{spec}}$, implying the role of the clear Ly$\alpha$ emission line in constraining the $z_{\mathrm{spec}}$. These quasars have $M_{1450}=-23.0$ to -25.2 mag at $z_{\mathrm{spec}}=4.7$ to $5.6$. 

\indent In addition to the best-fit $z_{\mathrm{spec}}$, we calculated the $z_{\mathrm{Ly}\alpha}$ by comparing the location of the strongest point in the Ly$\alpha$ emission line to the rest wavelength of the line (1216 \r{A}). Although the blueward of the Ly$\alpha$ emission line was attenuated by the neutral hydrogen in the IGM, the $z_{\mathrm{Ly}\alpha}$ is still consistent with the locations of the strong emission lines such as Ly$\alpha$ and C \Romannum{4}, especially for HSC J233107-001014. The $z_{\mathrm{Ly}\alpha}$ values are also provided in Table~\ref{tab:specfit}.  

\begin{figure}[t]
\centering
\includegraphics[width=80mm]{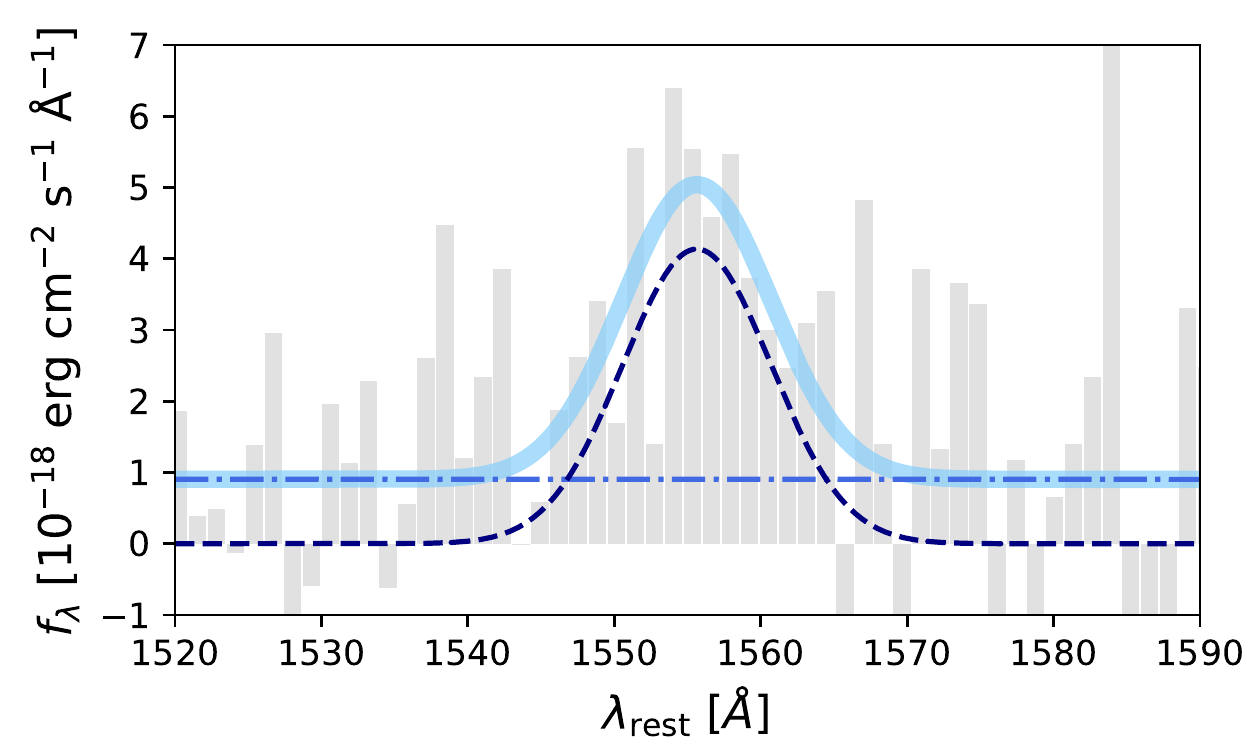}
\caption{Spectral modeling for the C\Romannum{4} emission line. We plot the binned spectrum of HSC J233107-001014 in the gray bar. The thick line indicates the best-fit model which is the summation of the power-law continuum and C\Romannum{4} emission line modeled as a single Gaussian distribution. The former and latter is plotted in the dot-dashed line and the dashed line, respectively. \label{fig:C4line}}
\end{figure}

\begin{figure}[t!]
\centering
\includegraphics[width=80mm]{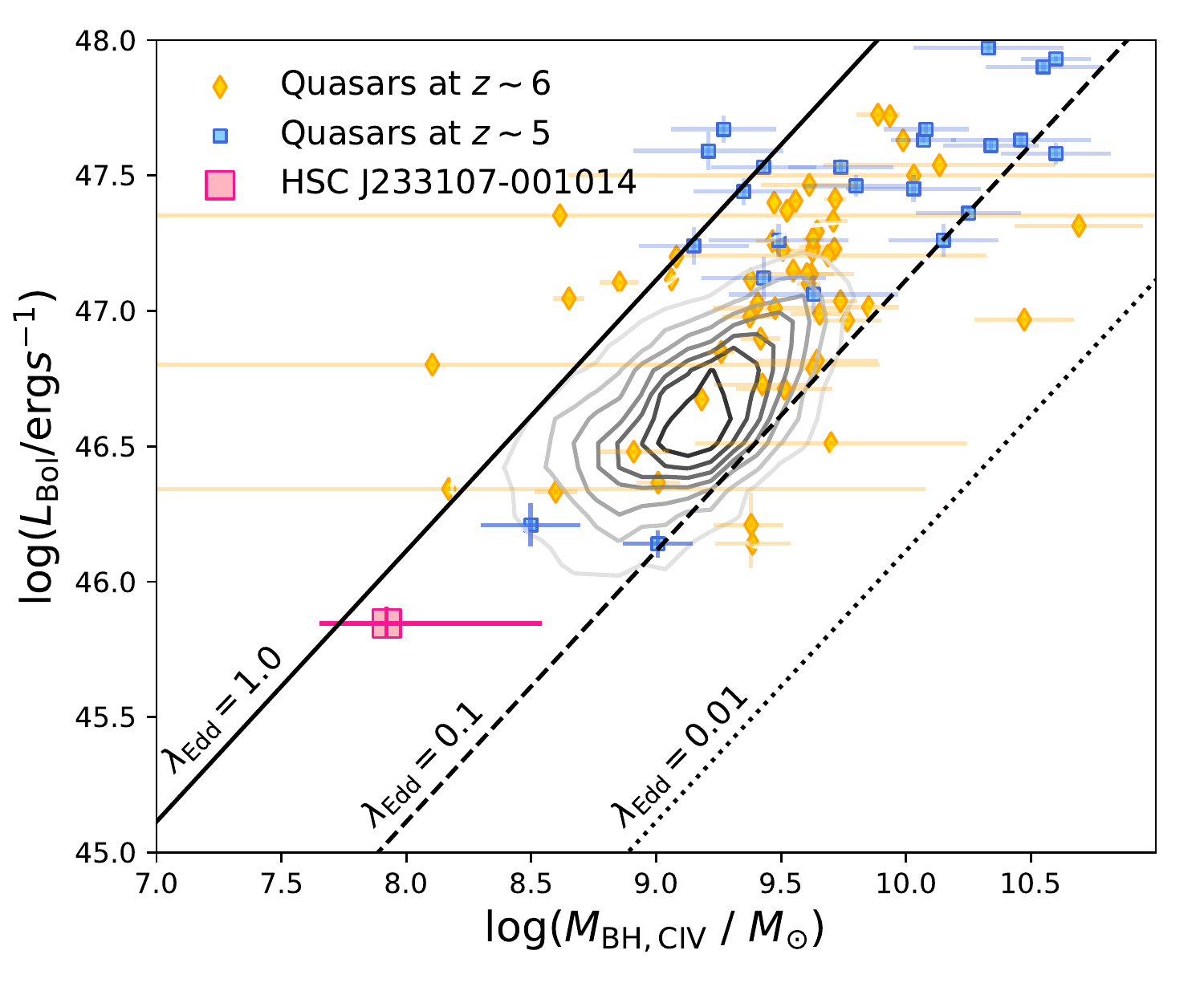}
\caption{$M_{\mathrm{BH}}$-$L_{\mathrm{Bol}}$ distributions of quasars, from $M_{\mathrm{BH}}$ measurements based on C\Romannum{4} emission line. The contours show the locations of the quasars at $z\sim2$ on the $M_{\mathrm{BH}}$-$L_{\mathrm{Bol}}$ plane \citep{Shen+2011}. The blue squares represent quasars at $z\sim5$ \citep{Jun+2015, Ikeda+2017}, while the gold diamonds indicate quasars at $z\sim6$ \citep{Jiang+2007, KimYJ+2018, Shen+2019}. Our faint quasar at $z\sim5.0$ is marked with the pink square. The solid, dashed, and dotted lines in the panels are corresponds to $\lambda_{\mathrm{Edd}}=1, 0.1$, and $0.01$, respectively.\label{fig:MBHvsLBol}}
\end{figure}

\subsection{BH mass and Eddington ratio}
\indent One interesting question is whether high-redshift quasars are more vigorously growing than lower-redshift quasars \citep{Willott+2010, KimYJ+2018, Onoue+2019, Shen+2019}. Recent studies of quasars at $z\gtrsim6$ have found that SMBHs in the bright quasars are vigorously growing with the Eddington ratios ($\lambda_{\mathrm{Edd}}$) of $\lambda_{\mathrm{Edd}} \sim 1$ \citep{Mortlock+2011, Banados+2018, Yang+2020a}. On the other hand, other studies find that fainter quasars at $z>6$ have $\lambda_{\mathrm{Edd}} \sim 0.1$, on par with quasars at $z=2$ to 3 \citep{Shen+2011, Mazzucchelli+2017, KimYJ+2018, Onoue+2019, Shen+2019}. Here, we investigate the supermassive black hole mass ($M_{\mathrm{BH}}$) and $\lambda_{\mathrm{Edd}}$ of one of our sample, HSC J233107-001014 for which C \Romannum{4} is detected.

\indent First, we transferred the observed frame to rest frame by using $z_{\rm spec}$. Then, we modeled continuum using a power-law function ($f_{\lambda} = \zeta \lambda^{\alpha_{\lambda}}$). Although the continuum consists of not only the power-law continuum but also Fe \Romannum{2} complex and Balmer continuum, we considered the former only due to the lack of sensitivity in its overall spectrum and an insignificant influence of the Fe \Romannum{2} emissions on the C \Romannum{4} line properties (e.g., \citealt{Shen+2011}). We fitted the power-law continuum to the fluxes of which the wavelength range is not close to those of the broad emission lines (e.g., Ly$\alpha$, C \Romannum{4}). To prevent large uncertainty of $\alpha_{\lambda}$ caused by poor SNR $\sim 1-3$ of the binned spectrum, it was inevitable to use global parts of the spectrum corresponding to $\lambda = 1260 - 1510$ and $1580 - 1660$, and assume $\zeta = 1$. The fitted power-law continuum was subtracted from the spectrum. 

\indent We also modeled the continuum-subtracted C \Romannum{4} emission line with a single Gaussian profile. Even though the C \Romannum{4} line is frequently regarded as the sum of multiple Gaussian profiles (e.g., \citealp{Jun+2015, Zuo+2020}), we did not include additional Gaussian components because the spectrum has a SNR too low to discern multiple components. To estimate the FWHM of the C \Romannum{4} emission line ($\mathrm{FWHM}_{\mathrm{C\Romannum{4}}}$) and its error, we generated 10,000 mock spectra by assuming the probability density function of the flux follows a Gaussian distribution. We calculated an $\mathrm{FWHM_{C\Romannum{4}, init}}$ for each spectrum, then the $\mathrm{FWHM}$ of our instrument ($\mathrm{FWHM_{int}} \sim 230$ km~s$^{-1}$) was subtracted from the estimated FWHMs, resulting in $\mathrm{FWHM_{C\Romannum{4}}}$ = $\sqrt{\mathrm{FWHM_{C\Romannum{4}, init}^{2} - \mathrm{FWHM_{int}}^{2}}}$. Adopting 16 and 84 percentiles of the $\mathrm{FWHM_{C\Romannum{4}}}$ histogram as 1-$\sigma$ uncertainties, the $\mathrm{FWHM}_{\mathrm{C\Romannum{4}}}$ is $2230_{-640}^{+1560}$ km~s$^{-1}$. The spectral fitting result of the C \Romannum{4} line is shown in Figure~\ref{fig:C4line}. 

\indent The mass scaling relation based on the C \Romannum{4} line \citep{Vestergaard+2006} is expressed as,

\begin{equation}
M_{\mathrm{BH}} = {\bigg[\frac{\mathrm{FWHM}_{\mathrm{C\Romannum{4}}}}{1000 \mathrm{km}\,\mathrm{s}^{-1}}\bigg]}^{2} {\bigg[\frac{\lambda L_{\lambda}( 1350 ~\mathrm{\r{A}})}{10^{44} \mathrm{erg} s^{-1}}\bigg]}^{0.53} \times 10^{6.6} M_{\odot}.
\end{equation}
$L_{\lambda}$(1350 \r{A}) is calculated from the best-fit SED model for the spectrum of HSC J233107-001014. The resulting $\log(M_{\mathrm{BH}}/M_{\mathrm{sun}})$ is $7.92_{-0.27}^{+0.62}$ for the virial factor of 5.1 \citep{Woo+2013}. 

\indent We caution that C \Romannum{4}-based $M_{\mathrm{BH}}$ could be systematically biased than the Balmer line-based $M_{\mathrm{BH}}$, since the C \Romannum{4} line profile could be seriously affected by non-virial motion (e.g., \citealt{Sulentic+2017}). To correct for systematic bias, some studies have considered the relation between physical parameters (e.g., C \Romannum{4} blueshift, Eddington ratio, C \Romannum{4} line asymmetry, and so on) and the difference between C \Romannum{4}-based and Balmer-based BH masses \citep{Coatman+2017, Marziani+2019, Zuo+2020}. Although the corrected $\mathrm{FWHM}_{\mathrm{C\Romannum{4}}}$ based on the relations could improve the $M_{\mathrm{BH}}$ estimates, we did not apply the correction due to large uncertainties in the physical parameters caused by the low SNR of the spectra. 


\indent Using the bolometric correction derived in \citep{Runnoe+2012}, which is

\begin{equation}
\log (L_{\mathrm{Bol}}) = 4.745 + 0.910\,\log(\,1450~\mathrm{\r{A}} \,L_{\lambda}(1450~\mathrm{\r{A}})\,),
\end{equation}

\noindent we calculated the bolometric luminosity of the quasar. For computing the Eddington luminosity $L_{\mathrm{Edd}}$, we use the following equation,

\begin{equation}
\log (L_{\mathrm{Edd}}) = 1.3 \times 10^{38} \times M_{\mathrm{BH}}/M_{\odot}\,\, \mathrm{erg} s^{-1}.
\end{equation}
 
The Eddington ratio is $L_{\mathrm{Bol}}/L_{\mathrm{Edd}} = 0.64_{-0.41}^{+0.93}$. Table~\ref{tab:Mbhinfo} provides the properties of HSC J233107-001014 derived from the line fitting. Figure~\ref{fig:MBHvsLBol} compares Eddington ratios of quasars at different redshifts in a $M_{\mathrm{BH}}$ versus $L_{\mathrm{Bol}}$ plane. Compared to quasars at similar luminosities ($L_{\mathrm Bol} \sim 10^{46}$ erg s$^{-1}$; \citealp{Shen+2011, Trakhtenbrot+2011, Ikeda+2017}), this quasar has somewhat a large $\lambda_{\mathrm{Edd}}$ if not exceptional. On the other hand, this $\lambda_{\mathrm{Edd}}$ is on par with brighter quasars at $z\sim5$ (e.g., \citealp{Trakhtenbrot+2011, Ikeda+2017, Jeon+2017}). This implies that not every faint quasar has smaller $\lambda_{\mathrm{Edd}}$ compared to the bright one. However, more faint quasars with well-measured BH properties should be required to better understand the accretion activities of $z\sim5$ quasars.

\begin{figure}[t!]
\centering
\includegraphics[width=80mm]{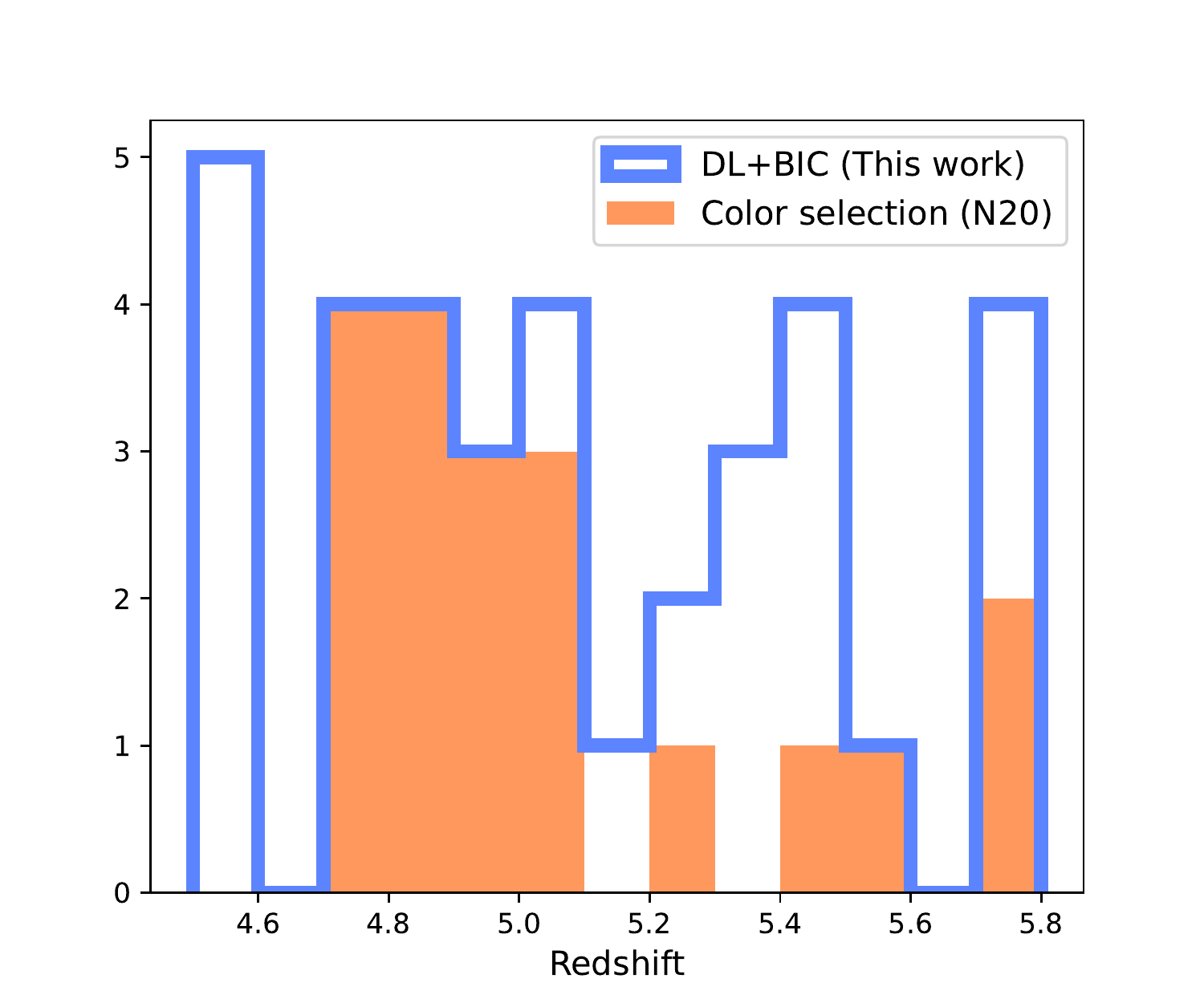}
\caption{Redshift histograms of the final quasar candidates (blue) and the final quasar candidates satisfying traditional color selection (\citetalias{Niida+2020}, orange). \label{fig:zhist}}
\end{figure}

\begin{table}
\caption{\label{tab:Mbhinfo}}
\centering
\begin{tabular}{lr}
\toprule 
HSC J233107-001014 & \\
\midrule
$\mathrm{FWHM_{C\Romannum{4}}}$ [km~s$^{-1}$] & $2230_{-640}^{+1560}$ \\
log$(L_{\mathrm{Bol}}\,[\mathrm{erg}\mathrm{s}^{-1}])$ & $45.85_{-0.05}^{+0.06}$\\
log$(M_{\mathrm{BH}}\,[M_{\odot}]$) & $7.92_{-0.27}^{+0.62}$ \\
$\lambda_{\mathrm{Edd}}$ & $0.64_{-0.41}^{+0.93}$\\
\bottomrule
\end{tabular}
\end{table}

\section{Implications on quasar selection \label{sec:efficiency}}
The confirmation of 4 new quasars brings the total number of spectroscopically confirmed quasars to 9 among 35 candidates. None of the candidates have been shown to be non-quasars so far, suggesting a very high confirmation rate. In addition, there is one quasar candidate with medium-band data, which can be almost certain to be a quasar \citep{Shin+2020}. 

\indent To quantitatively evaluate the effectiveness of our selection, we compare it to the traditional color selection of \citetalias{Niida+2020}. We checked whether the 35 final candidates can be selected by \citetalias{Niida+2020} selection. It misses 16/35 ($\sim 46 \%$) of the final candidates. Figure~\ref{fig:zhist} shows redshift histograms of the final candidates and the final candidates that meet \citetalias{Niida+2020} selection. Since the \citetalias{Niida+2020} selection has a high completeness value only when searching for quasars at $4.7 < z < 5.1$, it is difficult to select quasars at $z < 4.7$ or $z > 5.1$. However, our selection method can find quasar candidates at a broader redshift range than the \citetalias{Niida+2020} selection does, allowing us to increase the number of quasar sample at $z\sim5$. 

\indent We also calculate the recovery rate of known quasars at $4.5 < z < 5.5$. Our selection process recovers 5/6 known quasars and 4/4 newly discovered quasars (= 9/10, $\sim 90 \%$). We find that one quasar that has been missed is in the parameter space where the completeness is low (see \citealp{Shin+2022}, accepted), suggesting that our quasar recovery rate is as high as expected. To firmly confirm the effectiveness of our selection, spectroscopic observations for the 26 remaining candidates is required.

\section{Summary\label{sec:sum}}
We performed spectroscopic observations for four candidates with $M_{1450} \gtrsim -25.0$ mag at $z\sim5$ utilizing the DBSP on the 200-inch Hale Telescope in the Palomar observatory. The candidates were selected by deep learning and Bayesian information criterion (\citealp{Shin+2022}, accepted). Each candidate has a strong Ly$\alpha$ emission line and clear break near the line in its spectrum, suggesting all the candidates are quasars at $z\sim5$. 4$/$4 spectroscopic confirmation rate implies the validity of our novel selection approach. Our selection method provides a possible way for efficiently selecting high-redshift quasars at unbiased redshift ranges from future surveys.

\indent Since HSC J233107-001014 has a strong C \Romannum{4} emission line as well among the quasars, we calculated the marginal BH mass for the quasar, resulting in $10^{8} M_{\odot}$. The $\lambda_{\mathrm{Edd}}$ of the quasar is $\sim 0.6$, although most quasars with a similar luminosity ($L_{\mathrm{Bol}} \sim 10^{46}$ erg$s^{-1}$) to the quasar have lower Eddington ratios. To better understand the early growth of SMBHs, more faint quasars with $L_{\mathrm{Bol}} \lesssim 10^{46}$ erg$s^{-1}$ should be investigated. 

\acknowledgments
This research was supported by the National Research Foundation of Korea (NRF) grants No. 2020R1A2C3011091 and No. 2021M3F7A1084525, funded by the Ministry of Science and ICT (MSIT). S. S. acknowledges the support from the Basic Science Research Program through the NRF funded by the Ministry of Education (No. 2020R1A6A3A13069198). Y. K. was supported by the NRF grant funded by the MSIT (No. 2021R1C1C2091550). He acknowledges the support from the China Postdoc Science General (2020M670022) and Special (2020T130018) Grants funded by the China Postdoctoral Science Foundation. This research uses data obtained through the Telescope Access Program (TAP) (PID: CTAP2020-B0043 and CTAP2021-A0032), which has been funded by the National Astronomical Observatories of China, the Chinese Academy of Sciences, and the Special Fund for Astronomy from the Ministry of Finance. Observations obtained with the Hale Telescope at Palomar Observatory were obtained as part of an agreement between the National Astronomical Observations, Chinese Academy of Sciences, and the California Institute of Technology.

\indent The Hyper Suprime-Cam (HSC) collaboration includes the astronomical communities of Japan and Taiwan, and Princeton University. The HSC instrumentation and software were developed by the National Astronomical Observatory of Japan (NAOJ), the Kavli Institute for the Physics and Mathematics of the Universe (Kavli IPMU), the University of Tokyo, the High Energy Accelerator Research Organization (KEK), the Academia Sinica Institute for Astronomy and Astrophysics in Taiwan (ASIAA), and Princeton University. Funding was contributed by the FIRST program from the Japanese Cabinet Office, the Ministry of Education, Culture, Sports, Science and Technology (MEXT), the Japan Society for the Promotion of Science (JSPS), Japan Science and Technology Agency (JST), the Toray Science Foundation, NAOJ, Kavli IPMU, KEK, ASIAA, and Princeton University. 

\indent This paper makes use of software developed for the Large Synoptic Survey Telescope. We thank the LSST Project for making their code available as free software at  http://dm.lsst.org

\indent This paper is based on data collected at the Subaru Telescope and retrieved from the HSC data archive system, which is operated by the Subaru Telescope and Astronomy Data Center (ADC) at National Astronomical Observatory of Japan. Data analysis was in part carried out with the cooperation of Center for Computational Astrophysics (CfCA), National Astronomical Observatory of Japan. The Subaru Telescope is honored and grateful for the opportunity of observing the Universe from Maunakea, which has the cultural, historical and natural significance in Hawaii.


\end{document}